\documentclass[aps,prl,reprint,superscriptaddress, nobibnotes, nofootinbib]{revtex4-1}

\usepackage{graphicx}
\usepackage{color}
\usepackage[dvipsnames]{xcolor}
\usepackage[normalem]{ulem}

\newcommand{\blfootnote}[1]{
  \begingroup
  \renewcommand\thefootnote{}\footnote{#1}
  \addtocounter{footnote}{-1}%
  \endgroup
}

\begin{document}

\title{Solid-state qubits integrated with superconducting through-silicon vias}

\author{D. R. W. Yost}
    \thanks{Authors contributed equally to this work}
    \affiliation{MIT Lincoln Laboratory, 244 Wood Street, Lexington, MA 02421}
\author{M. E. Schwartz}
    \thanks{Authors contributed equally to this work}
    \affiliation{MIT Lincoln Laboratory, 244 Wood Street, Lexington, MA 02421}
\author{J. Mallek}
    \thanks{Authors contributed equally to this work}
    \affiliation{MIT Lincoln Laboratory, 244 Wood Street, Lexington, MA 02421}
\author{D. Rosenberg}
    \affiliation{MIT Lincoln Laboratory, 244 Wood Street, Lexington, MA 02421}
\author{C. Stull}
    \affiliation{MIT Lincoln Laboratory, 244 Wood Street, Lexington, MA 02421}
\author{J. L. Yoder}
    \affiliation{MIT Lincoln Laboratory, 244 Wood Street, Lexington, MA 02421}
\author{G. Calusine}
    \affiliation{MIT Lincoln Laboratory, 244 Wood Street, Lexington, MA 02421}
\author{M. Cook}
    \affiliation{MIT Lincoln Laboratory, 244 Wood Street, Lexington, MA 02421}
\author{R. Das}
    \affiliation{MIT Lincoln Laboratory, 244 Wood Street, Lexington, MA 02421}
\author{A. L. Day}
    \affiliation{MIT Lincoln Laboratory, 244 Wood Street, Lexington, MA 02421}
\author{E. B. Golden}
    \affiliation{MIT Lincoln Laboratory, 244 Wood Street, Lexington, MA 02421}
\author{D. K. Kim}
    \affiliation{MIT Lincoln Laboratory, 244 Wood Street, Lexington, MA 02421}
\author{A. Melville}
    \affiliation{MIT Lincoln Laboratory, 244 Wood Street, Lexington, MA 02421}
\author{B. M. Niedzielski}
    \affiliation{MIT Lincoln Laboratory, 244 Wood Street, Lexington, MA 02421}
\author{W. Woods}
    \affiliation{MIT Lincoln Laboratory, 244 Wood Street, Lexington, MA 02421}
\author{A. J. Kerman}
    \affiliation{MIT Lincoln Laboratory, 244 Wood Street, Lexington, MA 02421}
\author{W. D. Oliver}
 \affiliation{MIT Lincoln Laboratory, 244 Wood Street, Lexington, MA 02421}
 \affiliation{Research Laboratory of Electronics, Massachusetts Institute of Technology, 77 Massachusetts Avenue, Cambridge, MA 02139}

\date{\today}

\begin{abstract}
As superconducting qubit circuits become more complex, addressing a large array of qubits becomes a challenging engineering problem.
Dense arrays of qubits benefit from, and may require, access via the third dimension to alleviate interconnect crowding. 
Through-silicon vias (TSVs) represent a promising approach to three-dimensional (3D) integration in superconducting qubit arrays\textemdash provided they are compact enough to support densely-packed qubit systems without compromising qubit performance or low-loss signal and control routing.
In this work, we demonstrate the integration of superconducting, high-aspect ratio TSVs\textemdash 10 $\mu$m wide by 20 $\mu$m long by 200 $\mu$m deep\textemdash with superconducting qubits.
We utilize TSVs for baseband control and high-fidelity microwave readout of qubits using a two-chip, bump-bonded architecture.
We also validate the fabrication of qubits directly upon the surface of a TSV-integrated chip.
These key 3D integration milestones pave the way for the control and readout of high-density superconducting qubit arrays using superconducting TSVs.

\end{abstract}

\blfootnote{For correspondence: yost@ll.mit.edu; william.oliver@mit.edu}

\maketitle

\section{Introduction}

\begin{figure*}[ht]
\includegraphics{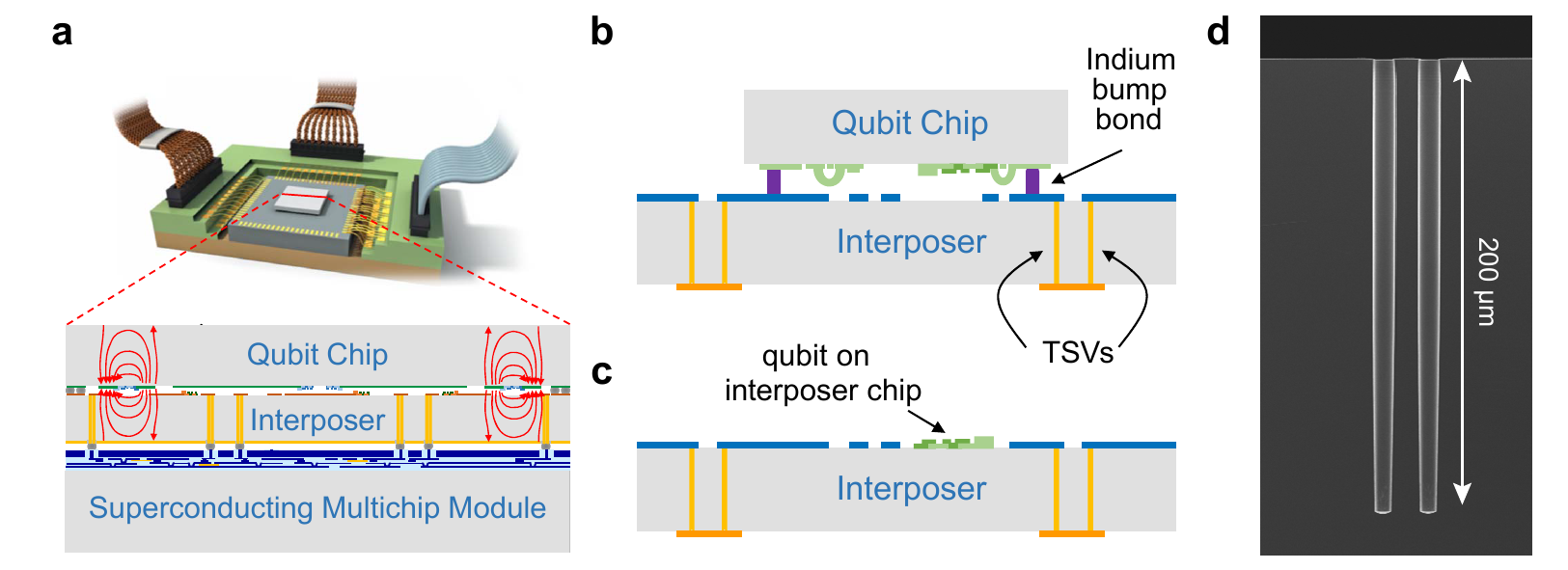}
\caption{\label{fig:ThreeStack}  3D-integration scheme for readout and control of qubit arrays.
\textbf{a} Schematic rendering of the ``3-stack'': a qubit chip (top) is connected to a wiring and signal processing circuitry chip (bottom) through an interposer chip (middle).
Superconducting through silicon vias (TSVs) provide electrical connection between the qubit and routing chips. Panel adapted from Ref. \cite{Rosenberg_2017}.
\textbf{b} ``2-stack'' in which qubits are biased and measured through TSVs on the interposer chip.
\textbf{c} Qubits fabricated directly on the surface of the interposer chip.
\textbf{d} SEM cross-section of an etched TSV before wafer thinning, demonstrating the high aspect ratio etch processes leveraged in this work. }
\end{figure*}

Superconducting qubits are lithographically defined electrical circuits comprising Josephson junctions, inductors, capacitors, and interconnects that are engineered to behave as well-isolated quantum-mechanical two-level systems \cite{Blais_2004, Koch_2007, Krantz_2019}.
Superconducting qubits use materials such as aluminum (Al), titanium nitride (TiN) and niobium (Nb) and substrates such as silicon (Si) and sapphire  (Al$_2$O$_3$) that are compatible with\textemdash and strongly leverage\textemdash CMOS manufacturing toolsets and approaches \cite{Oliver_2013}.
Their lithographic scalability, compatibility with microwave control, and operability at the nanosecond timescale have all converged to place superconducting qubits at the forefront of nascent quantum processor development.
Current state-of-the-art superconducting circuits feature single-qubit gate fidelities exceeding 99.9\% and two-qubit gate fidelities exceeding 99\% \cite{Kjaergaard_2019}.

To date, most multi-qubit demonstrations on superconducting qubit processors have been performed in a planar architecture: qubits and interconnects have been co-fabricated on a single chip surface, with control and readout wiring brought in laterally from the edges of the chip \cite{otterbach_2017, roushan_2017, havlicek_2019}.
While this approach can potentially support tens of qubits, interconnect crowding precludes lateral interconnection as circuits become larger.

Addressing the interior elements of a large two-dimensional (2D) array requires moving out of a planar geometry to route wires past the outer elements of the array.
In the short- and medium-term, accessing the interior of a qubit array can be achieved with a single extra metallization layer (i.e. air bridges and standard flip-chip bonding) or by a single layer of vertical I/O (i.e. pin-chips, pogo pins, and similar technologies) \cite{Rosenberg_2019, Foxen_2017, Chen_2014, Rosenberg_2017, Niedzielski_2019, Bronn_2018, Mariantoni_2018, Arute_2019}.
However, larger and more complex quantum system architectures may need to utilize multiple levels of qubits and complex signal routing, which necessitates the development of multi-layer control and routing capabilities.
In addition, as processor size increases, through-substrate ground connections will need to be introduced to eliminate parasitic chip modes, in the same spirit as via walls in large interposer boards for classical electronics.
All of these considerations point to a need for more sophisticated and capable 3D integration approaches for superconducting qubits.

3D integration as developed for conventional computing and imaging applications conceptually offers a solution for signal routing.
However, many pre-existing 3D integration techniques involve processes and materials that are incompatible with high-coherence superconducting qubits.
Many through-silicon via (TSV) interposer, multi-tier stacking, and semiconductor technologies use deposited inter-layer dielectrics (ILDs) to isolate signals in multi-layer wiring \cite{Gambino_2015, Iyer_2015}.
However, industry-standard ILD materials, such as silicon dioxide (SiO$_2$) and silicon nitride (Si$_3$N$_4$), decrease qubit lifetimes due to the interaction of the qubit electric field with the many defects present in such lossy, amorphous dielectrics \cite{Martinis_2005}.
In addition, common conductors used for metallization, such as copper, are not superconducting.
Electric currents traveling in such normal metals cause heating and excite quasiparticles that degrade qubit performance \cite{Patel_2017}.
Monolithic approaches using superconducting Nb circuitry with multiple wiring layers for dense signal routing can provide adequate connectivity, and even a pathway to integrated superconducting control electronics \cite{Tolpygo_2014}, yet these are also associated with low qubit coherence due to, for example, lossy integrated dielectric films \cite{Berns_2008, Oliver_2009, Johnson_2011, Harris_2018}.

To mitigate the incompatibility of both monolithic and standard 3D integration techniques with high-coherence qubits, we have developed a ``3-stack'' chip concept that enables vertical signal delivery to a 2D array of qubits while maintaining qubit coherence \cite{Rosenberg_2017, Niedzielski_2019}.
The 3-stack (Figure~\ref{fig:ThreeStack}a) consists of three chips, fabricated independently according to best practices for each particular functionality, that are bump-bonded together using thermo-compression bonding of indium pillars.
The top chip\textemdash the qubit chip\textemdash contains high-coherence superconducting qubits fabricated on a silicon wafer.
The bottom chip\textemdash a superconducting multi-chip module (SMCM)\textemdash has multiple planarized superconducting layers connected by vias and uses processes that have been advanced by digital superconducting electronics \cite{Tolpygo_2014}.
The SMCM chip may be used for both passive signal routing and active elements for qubit control and measurement such as single-flux-quantum circuits \cite{li_2019} or travelling wave parametric amplifiers \cite{Macklin_2015}.

The middle chip\textemdash known as the interposer chip\textemdash is the key innovation in this work.
It contains TSVs lined with superconducting titanium nitride that connect the qubit and SMCM chips without resistive loss and without requiring close proximity between the qubits and the routing chip.
When paired with indium bump bonds, this enables a low-resistance electrical path between the three chips \cite{Rosenberg_2017, Foxen_2017}.
The TSV-integrated interposer makes it possible to leverage the multi-layer signal routing in the SMCM while maintaining isolation of qubits from lossy dielectrics.
In particular, the silicon that comprises the interposer chip both provides physical distance to prevent overlap between the qubit electrical field and that of the dielectrics in the SMCM, and spreads the qubit electromagnetic field over a large mode volume to reduce sensitivity to loss modes.
The interposer chip therefore alleviates a scaling bottleneck for the superconducting qubit platform.

In this paper, we present the integration of high aspect ratio superconducting TSVs with superconducting qubits.
Our fabrication toolset enables compact TSVs\textemdash with a footprint of 10~$\mu$m x 20~$\mu$m passing through a 200~$\mu$m chip\textemdash which allow for close packing of the TSVs and supports integration of TSVs directly into qubit readout and control circuitry.
We demonstrate these capabilities in a ``2-stack" that comprises the qubit chip and the TSV-integrated interposer chip.
The chips are mechanically and electrically integrated using indium bump-bonding \cite{Rosenberg_2017}, as shown in Figure~\ref{fig:ThreeStack}b and described in the Methods.
A TSV pitch of 1 TSV per 100~$\mu$m enables up to 10,000 TSV interconnects on a 10~mm x 10~mm chip, ensuring consistent ground connection across the chip; tighter pitches are also used in readout and control circuitry~(Figure~\ref{fig:ThreeStack}d).
We show that our process is compatible with high-performance qubits by demonstrating 10-20 $\mu$s $T_1$ relaxation times both for qubits separately fabricated on a qubit chip and bump-bonded to the interposer in the 2-stack, and for qubits that are fabricated directly on the surface of the interposer chip (Figure~\ref{fig:ThreeStack}c).
This work represents an important step towards realizing high-density 3D integration of arrays of qubits.

\section{Results}
We benchmark the integration of superconducting TSVs with superconducting qubits using capacitively-shunted flux qubits (CSFQs) \cite{Nakamura_1999, Yan_2016}.
The CSFQs used in this work are designed for quantum annealing applications requiring tunable magnetic fields, large qubit-to-qubit coupling, and high connectivity \cite{Weber_2017}.
Planar equivalents of these devices exhibit relaxation times of T$_1 = 10-20$~$\mu$s \cite{Rosenberg_2017}, much longer than those demonstrated in commercial monolithic quantum annealing systems \cite{Harris_2018}.
Control of annealing qubits requires several mA of baseband ($< 100$~MHz) current in order to tune the qubit Hamiltonian.
More generally, qubit excitation and readout requires the ability to send high-fidelity microwave signals in the 3-10~GHz range.
Here, we demonstrate such baseband and microwave control of CSFQs using TSV-integrated bias and readout elements.

\begin{figure}[ht]
\includegraphics[width=3.25in]{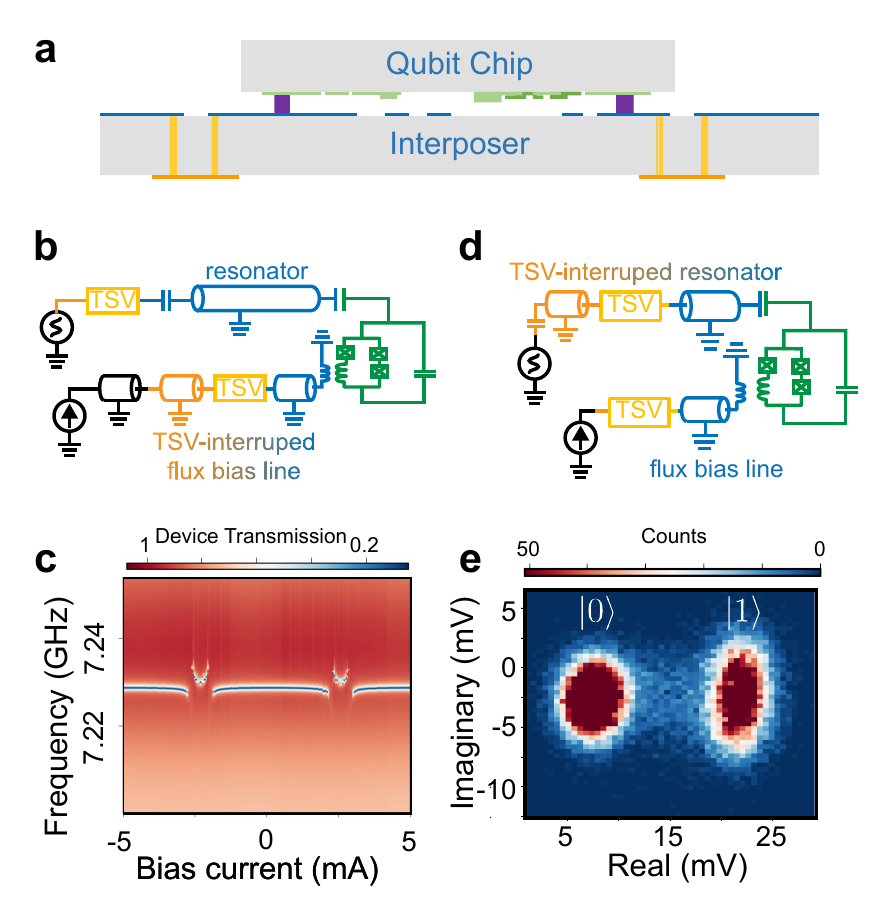}
\caption{
2-stack devices used to demonstrate control and readout circuitry incorporating TSVs.
\textbf{a} Schematic showing color coding. Green: qubit chip; blue: top of interposer chip; yellow: TSVs; orange: bottom of interposer chip; black: room temperature electronics.
Circuit schematics show the surface placement of qubits, resonators, and bias lines, including TSV transitions in the flux bias line (\textbf{b}) and readout resonator (\textbf{d}).
The device was probed from the top surface through an additional TSV transition.
\textbf{c} Spectroscopy of a readout resonator taken while biasing a qubit using a flux bias line incorporating a TSV transition.
Four avoided crossings between the qubit and the resonator are observed, indicating that the qubit has tuned through more than a full flux quantum.
\textbf{e} Single shot readout using a readout resonator incorporating a microwave-optimized TSV transition.
The data represent sequential preparation of the qubit in the $|0\rangle$ and $|1\rangle$ states, respectively as labeled.
The readout partially excites the $|1\rangle$ state into $|2\rangle$ due to the relatively high photon number used for readout \cite{Sank_2016}.
This leads to the slight skew in the distribution corresponding to $|1\rangle$, but does not impact the fidelity of the measurement because the $|0\rangle$ state is unaffected.}
\label{fig:two-stack}
\end{figure}

\subsection{Qubit control and readout through interposer chip}

Our first set of devices demonstrate the active control of a qubit using off-chip lines that pass from one side of the interposer to the other via a 50~$\Omega$ TSV transition (Figure~\ref{fig:two-stack}). Specifically, 2-stack devices with TSVs in flux bias lines, used for baseband qubit frequency control, and in coplanar waveguide (CPW) resonators, used for microwave qubit readout, were designed and fabricated.
Devices testing the baseband and microwave control functionality were separately fabricated and measured; the mean qubit lifetime across all such devices was T$_1 = 10~\mu$s (Table 1).

In Figure \ref{fig:two-stack}b-c, we demonstrate baseband control of the CSFQ frequency using a TSV-integrated flux bias line inductively coupled from the interposer chip to the qubit chip.
A direct current is introduced to the bottom side of the interposer chip through a CPW input line.
The current passes through a 50~$\Omega$ baseband transition\textemdash comprising a set of TSVs, surrounding TSV ground shields, and top- and bottom- side metallizations\textemdash that is designed to ensure good signal integrity across baseband frequencies.
This current continues through a second CPW section on the top side of the interposer chip, and then couples inductively to a CSFQ located on the qubit chip with a mutual inductance of $M = 0.4$ pH.
This suffices to tune the CSFQ's first transition frequency over a full flux quantum by applying approximately 5 mA to the bias line, as evidenced by the periodicity of the avoided crossing between the qubit and the resonator (Figure~\ref{fig:two-stack}c).
This is well within the bounds of the typical TSV critical current, which exceeds 10 mA.
We see no heating of the dilution refrigerator during the persistent application of this current, which is consistent with all aspects of the circuit remaining superconducting.
Thus, we demonstrate that the interposer and TSV-integrated baseband transition support CSFQ frequency tuning through several flux quanta.

We next use a TSV-integrated resonator as a test of the ability of the TSVs to carry microwave signals (Figure~\ref{fig:two-stack}d-e).
Splitting a resonator or other microwave component across the two interposer planes enables novel designs and a more compact per-qubit footprint.
In addition, from a technological perspective this test is particularly sensitive, because any signal distortions in the microwave range would impact our ability to read the qubit state with high fidelity.
We pattern the first half of a $\lambda/4$ CPW resonator onto the bottom of the interposer (Figure~\ref{fig:two-stack}d) and pass to the top of the chip using a 50~$\Omega$ TSV transition, this time optimized to maintain signal integrity at microwave frequencies (see Ref. \cite{Rosenberg_2019} for more details).
The second half of the resonator is patterned on the top of the interposer, where it capacitively couples to the CSFQ on the qubit chip.
Figure~\ref{fig:two-stack}e shows single-shot readout statistics for the CSFQ prepared in its ground (left) and excited (right) states, using a measurement integration time of 2.5 $\mu$s.
These data correspond to a separation infidelity of $\epsilon_S = 2.1 \times 10^{-5}$ (defined as the integrated overlap of the Gaussian distributions to which these data were fit).
Thus, we demonstrate that TSV transitions can support microwave signals sufficient to perform high-fidelity qubit readout.

\begin{figure}[hb]
\includegraphics[width=3.25in]{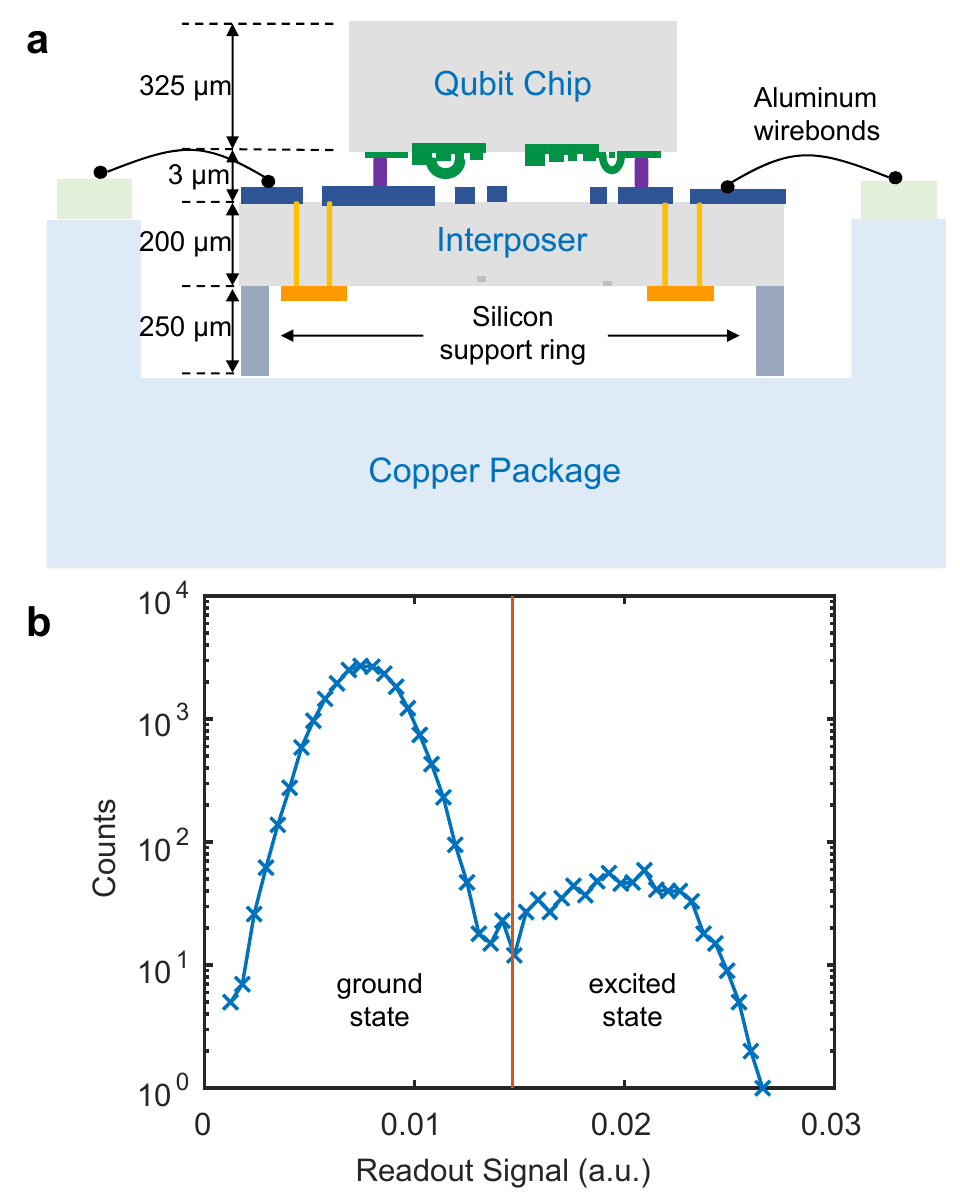}
\caption{\label{fig:thermalization}  Qubit thermalization through interposer chip.
\textbf{a} Schematic showing the path from the qubit chip to the thermal bath (i.e. the copper package).
The qubit chip thermalizes via superconducting indium (purple), aluminum (green), and titanium nitride (blue, orange) metallization layers and wirebonds, as well as through bulk silicon.
\textbf{b} Histogram of 21,000 single-shot measurements of a sample CSFQ that has been passively prepared in its thermal state. We extract an excited state probability of $P_e = 3.3\%$, which corresponds to an effective qubit temperature of $T_q = 52.5$ mK for a qubit with transition frequency $\omega_q/2\pi = 3.718$ GHz.}
\end{figure}

\subsection{Qubit thermalization in multi-chip architectures}

Strong thermalization in any multi-tier qubit stack is critical to ensure that qubits maintain the tens-of-mK temperatures required to avoid spurious excitations.
As shown schematically in Figure~\ref{fig:thermalization}a, the qubit chip thermalizes through its superconducting (and therefore poorly thermally conducting) indium bump bonds to the interposer chip.
The interposer chip is itself separated from the copper package\textemdash which provides the cold thermal bath\textemdash by a silicon support ring that mimics the SMCM chip and by superconducting aluminum wirebonds.
None of these are canonically good thermal conduction paths, so we must verify that qubits are able to sufficiently thermalize in the 2- and 3-stack architectures.

The ability to do high-fidelity single-shot readout using TSV connections enables us to probe qubit thermalization in the 2-stack architecture.
This can be tested with the same device used in Figure~\ref{fig:two-stack}d-e.
In Figure~\ref{fig:thermalization}b, we show a histogram of 21,000 single-shot measurements of a qubit that has been passively prepared in its thermal steady state.
A DC current of 2.3 mA was applied during the duration of this experiment; no heating of the dilution refrigerator was observed.
We measure an excited state probability $P_e = 3.3\%$, corresponding to an effective qubit temperature of $T_Q = 52.5$ mK at the qubit transition frequency 3.718 GHz.
This compares favorably to single-qubit-chip thermalization temperatures of 35-40 mK \cite{Jin_2015}.
Well-established techniques exist for initializing a pure quantum state at these temperatures \cite{johnson_2012, Riste_2012, Geerlings_2013}; we therefore confirm that qubit performance in the 2-stack is on par with that in single-chip architectures.

\subsection{Integration of qubits on TSV interposer surfaces}

\begin{figure*}[ht]
\includegraphics[width=7in]{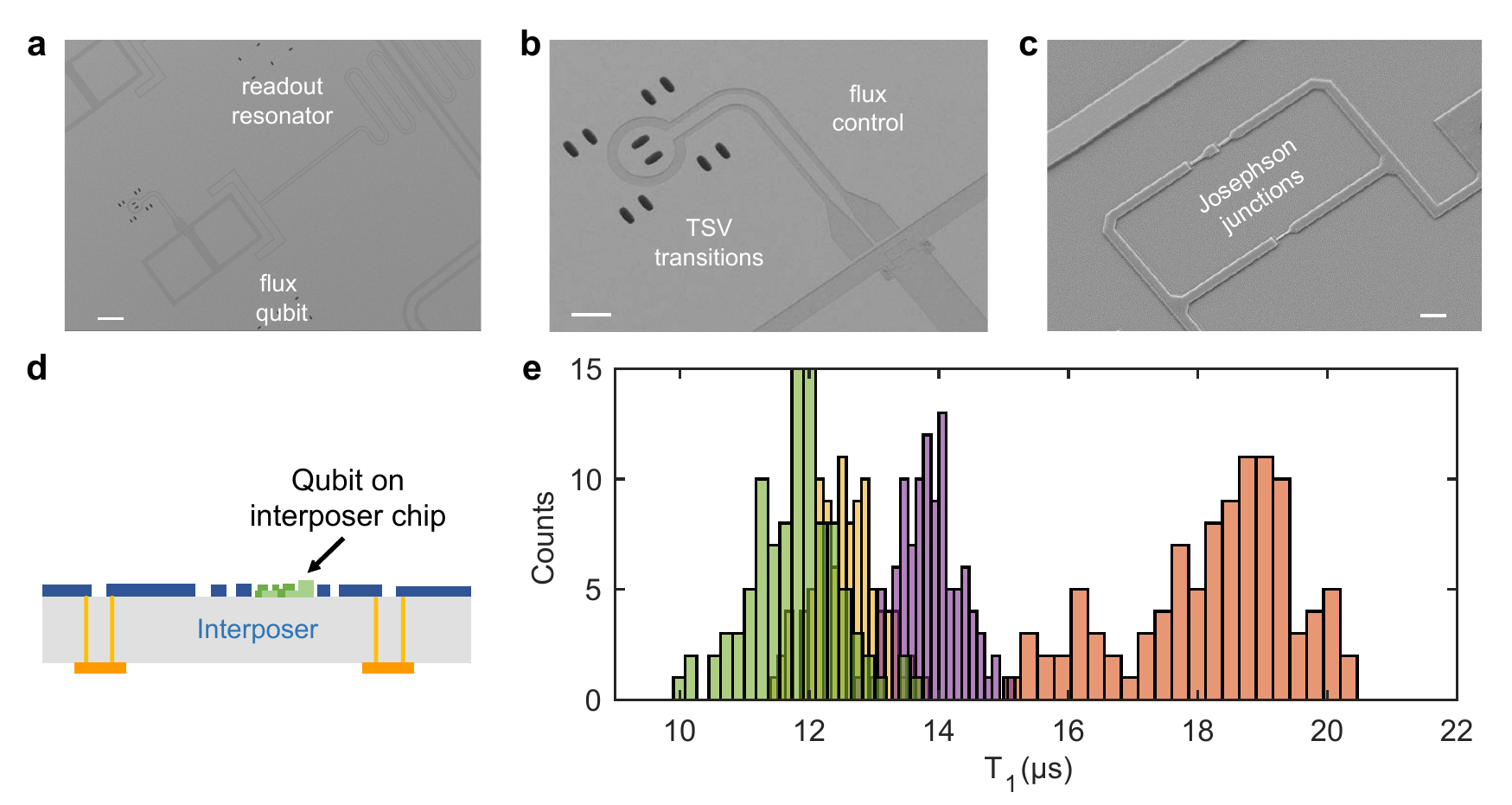}
\caption{Qubits fabricated directly on the surface of the TSV-integrated interposer chip
\textbf{a-c} SEM images of TSV-integrated qubits.
\textbf{a} A zoom-out showing the qubit, the readout resonator, and the TSV-integrated flux bias line.
Scale bar is 100 $\mu$m.
\textbf{b} A zoom-in showing the TSV flux line transition.
Scale bar is 30 $\mu$m.
\textbf{c} The three Josephson junctions that form the CSFQ loop.
Scale bar is 2 $\mu$m.
\textbf{d} Schematic of qubits patterned directly on the qubit-chip-facing surface of the interposer chip.
\textbf{e} A histogram of qubit relaxation times (T$_1$) measured for four qubits.
The mean T$_1$ across all four qubits is 12.5 $\mu$s.}
\label{fig:qubit_on_interposer}
\end{figure*}

To take full advantage of the additional surface that the top of the interposer chip provides, the interposer top should ideally be compatible with active Josephson junction elements including tunable couplers and qubits.
This is a highly nontrivial engineering challenge: the interposer chip top sees substantial mechanical and chemical processing far beyond that used for a standard qubit chip (see Methods).
Deposition and patterning processes all must be optimized for compatibility with these additional processing steps, imposing significant constraints.
Overcoming these challenges and adding coherent active elements and qubits on the interposer, however, pays dividends in opening up new architectural possibilities for denser qubit packing and novel qubit coupling schemes \cite{Weber_2017}.

In Figure~\ref{fig:qubit_on_interposer}, we demonstrate the fabrication of coherent CSFQs directly on the surface of the interposer chip. Figure~\ref{fig:qubit_on_interposer}a-c show scanning electron microscope (SEM) images of the top side of the interposer, including TSVs, control and readout circuitry, 
and Josephson junctions. 
Despite the process integration challenges required to create the TSVs, we measure relaxation times (T$_1$) for qubits fabricated on the interposer surface averaging 12.5 $\mu$s across four co-fabricated qubits (Figure~\ref{fig:qubit_on_interposer}e).
These coherence times are consistent with similar annealing qubits fabricated on silicon surfaces that have not undergone TSV processing steps \cite{Rosenberg_2017}.
Such qubits are sufficient for off-chip coupling elements for gate-model quantum computing and are on par with devices being developed for high-coherence annealing \cite{Rosenberg_2017, Quintana_2017}.
This validates an important new capability for the 2-stack and 3-stack architectures \cite{Yan_2018, Weber_2017}.

\section{Discussion}
In this work, we have demonstrated integration of superconducting TSVs into superconducting qubit readout and control circuitry, enabled by the small footprint and high aspect ratio of our superconducting TSVs.
The TSVs provide densely-packed ground connections across the interposer and enable active readout and control modalities to cross through the interposer without degrading signal performance and with the potential for reduced crosstalk.
The TSVs occupy a footprint that is orders of magnitude smaller than that of current-generation superconducting qubits, enabling the use of TSV-integrated control and readout elements without impacting the effective footprint of each qubit.
We have designed and demonstrated compact, 50~$\Omega$ TSV transitions that support both baseband and microwave signals, showing that we can route high-fidelity control and readout signals through the third dimension in a multi-chip stackup.
We show that qubits thermalize effectively in our multi-chip stackup, despite the additional thermal resistance introduced by the integration of multiple chips.
These capabilities confirm and demonstrate the utility of our TSV design and fabrication in enabling the readout and control of 3D integrated qubit arrays.

The utilization of TSV-integrated interposers demonstrated here advances the state of the art in superconducting TSVs for quantum computing applications in several important ways.
TSVs with sloped-wall geometries have been previously demonstrated to suppress substrate modes on qubit chips \cite{vahidpour_2017}; however, these TSVs are the same size as typical superconducting qubits ($\sim$~100-300~$\mu$m per side) and if incorporated into both readout and control circuitry would significantly increase the per-qubit footprint.
High-aspect ratio TSVs have been used to connect to large area kilopixel transition-edge sensor bolometer arrays \cite{Jhabvala_2014}; however, these TSVs lacked critical currents sufficient to enable flux control of qubit frequencies, and they required the use of atomic layer deposition instead of the chemical vapor deposition (CVD) used here.
The TSVs integrated into the interposer chip in this work combine the aspect ratio and the critical current necessary for full qubit readout and control.

Complex quantum system architectures may benefit from multiple levels of qubits and several routing layers.
Our ability to fabricate coherent CSFQs directly on the surface of interposer chips supports a variety of 3D-integrated qubit architectures.
One can use the top side of the interposer chip to introduce a second qubit plane and increase the qubit density beyond what is achievable in a single-tier architecture.
For applications in which different Josephson junction critical current densities are required for different circuit components those components can be fabricated on separate chips, thereby minimizing additional processing on each chip and increasing design flexibility.
Active elements, such as tunable couplers, can be fabricated on the surface of the interposer chip in order to enable novel schemes with stronger, more compact coupling \cite{Weber_2017}.
These 2-stack demonstrations passing high-fidelity baseband and microwave signals through compact TSVs are critical enabling capabilities for connecting superconducting multi-layer routing to high-coherence qubits. 
Future work will demonstrate the full 3-stack integration of qubits, interposer, and multilayer signal routing.

\section{Methods}

\begin{table*}[ht]
\begin{tabular}{|c|c|c|c|c|c|c|}
    \hline
    \multicolumn{7}{|c|}{Qubit Parameters} \\
    \hline
    Qubit & Qubit location & Qubit Freq. (GHz) & Resonator Freq. (GHz) & T$_1$ ($\mu$s) & T$_{2, Ramsey}$ $(\mu s)$ & T$_{2, Echo}$ $(\mu s)$  \\
    \hline
    a & Qubit chip & 4.189 & 7.165 & $10 \pm 3$ & n/m & n/m \\
    \hline
    b & Qubit chip & 4.091 & 7.229 & $9.1 \pm 0.8$ & $7.7 \pm 0.6$ \ & $8.5 \pm 0.6$ \\
    \hline
    c & Qubit chip & 3.671 & 7.259 & $10 \pm 3$ & n/m & n/m  \\
    \hline
    d & Qubit chip & 3.698 & 7.362 & $9 \pm 3$ & n/m & n/m  \\
    \hline
    e & Qubit chip & 3.718 & 5.281 & $9 \pm 2$ & n/m & n/m\\
    \hline
    f & Interposer & 3.359 & 7.324 & $18 \pm 1$ & $19 \pm 2$ & $27 \pm 2$  \\
    \hline
    g & Interposer & 2.483 & 7.395 & $12.6 \pm 0.5$ & $1.7 \pm 0.2$ & $1.6 \pm 0.2$  \\
    \hline
    h & Interposer & 3.788 & 7.469 & $13.8 \pm 0.5$ & $2.6 \pm 0.3$ & $3.3 \pm 0.5$\\
    \hline
    i & Interposer & 2.576 & 7.658 & $11.8 \pm 0.7$ & $3.0 \pm 0.4$ & $4.0 \pm 0.6$ \\
    \hline
    \hline
\end{tabular}
\caption{Qubit and resonator frequencies for all devices described in this paper, as well as T$_1$ relaxation times and T$_{2, Ramsey}$ and T$_{2, Echo}$ as measured at the flux-insensitive point. The notation ``n/m'' indicates quantities that were not measured.
Error bars represent the standard deviation of repeated measurements over several hours. The differences between the T$_1$ of the qubits on the qubit chip and on the interposer chip are not statistically significant.
The lifetime of flux qubits is typically limited by the size of the SQUID loop: larger loops have lower T$_1$, but are also amenable to stronger coupling to neighboring qubits, which has advantages in annealing applications.
Planar equivalents of the devices measured here (i.e. designed with identical SQUID loops and capacitances) typically have relaxation times of T1= 10-20 $\mu$s \cite{Rosenberg_2017}.
Because this study focused on T$_1$, the measurements of T$_2$ were typically taken from rough calibration scans and are underestimates of the true coherence of the devices.
Samples b and f were fine-tuned to extract the optimal phase coherence for those devices.}
\label{tab:quparams}
\end{table*}

\subsection{TSV-integrated interposer fabrication}
The fabrication of the TSV-integrated interposer follows a via-first process on an industry-standard 200~mm wafer fabrication line.
The TSVs are etched into 725~$\mu$m thick high resistivity silicon wafers.
A Bosch etch process consisting of rapidly alternating cycles of silicon etching and sidewall polymerization is used to etch high-aspect-ratio TSVs with smooth sidewalls (low scalloping).
We etch blind TSVs $>$200~$\mu$m deep with a sidewall slope $>89^\circ$.
A thin film of TiN is deposited using an optimized CVD process to provide superconducting film coverage along the sidewalls of the TSVs and over the wafer surface; this becomes the bottom of the wafer, as shown in orange in Figures~\ref{fig:ThreeStack}-\ref{fig:qubit_on_interposer}. This metal is subtractively patterned using thick resist and plasma etching.

Each interposer wafer is then flipped and bonded to a temporary carrier wafer using a Brewer wafer bond temporary adhesive.
The wafer is thinned by a combination of grinding and chemical mechanical polishing to reveal the opposite (top) side of the TSVs shown in blue in Figures~\ref{fig:ThreeStack}-\ref{fig:qubit_on_interposer}.
TiN is deposited on this newly revealed top surface using physical vapor deposition.
Interconnects, bias lines, resonators, and qubit capacitors are defined using subtractive etching.
Al/AlO$_{\text{x}}$/Al Josephson junctions for qubits and other active elements are fabricated using a Dolan bridge defined by electron-beam lithography and utilizing a double-angle shadow evaporation process \cite{Dolan_1977}.
Finally, the wafer is diced and released from the temporary carrier wafer.
Additional interposer fabrication details will be published in a forthcoming work by Mallek and Yost, \textit{et al}.

\subsection{Qubit chip fabrication}
Qubit chip devices are patterned on 50-mm-diameter high-resistivity ($>$10,000 $\Omega$-cm) silicon substrates.
The qubit devices consist of silicon hard-stop spacers \cite{Niedzielski_2019}, base metal, superconducting qubits, underbump metal, and indium bumps.
Base metallization components, including capacitive shunts, ground planes, and on-chip control and readout circuitry, are patterned by subtractive etching into aluminum.
The patterned surface is ion milled \emph{in situ} to remove native aluminum oxide prior to definition of qubit loops and Josephson junctions using double-angle shadow evaporation through Dolan-style bridges \cite{Dolan_1977}.
Indium bumps, which are $15~\mu$m in diameter and 5-15 $\mu$m tall, are patterned on the qubit chip on top of an underbump metallization layer of titanium/platinum/gold which ensures the indium makes good electrical contact with the underlying aluminum.
Additional fabrication details are available in Refs. \cite{Yan_2016, Rosenberg_2017}.

\subsection{Flip-chip integration}

Qubit and interposer chips are bump-bonded using thermocompression bonding.
For the chips used in this work, indium bumps are used only to connect the ground planes of the two chips; they do not carry active signals.
Bump bonding is performed at 105$^\circ$ C in a commercial thermocompression bonding system with a post-bond lateral accuracy of  $<1~\mu$m.
This alignment accuracy, which is verified with post-bond infrared imaging of alignment structures, has a negligible effect on the coupling between the qubits and elements on the interposer chip.
The post-bond vertical spacing between the two chips is 3~$\mu$m.

\subsection{Measurement}

The measurements in Figures \ref{fig:two-stack}-\ref{fig:qubit_on_interposer} are taken using standard dispersive measurement techniques \cite{Blais_2004}.
We couple a CSFQ biased at its the flux-insensitive point in the 2-5 GHz range to a CPW $\lambda/4$ resonator in the 7-8 GHz range.
The experiments were performed in a commercial BlueFors XLD-1000 dilution refrigerator operating at a base temperature of 12~mK.
The measurement chain incorporates a high-efficiency travelling-wave parametric amplifier that enables high-fidelity readout \cite{Macklin_2015}.
In Figure~\ref{fig:two-stack}a, we measure the transmission of a resonator that is capacitively coupled to a CSFQ in a 2-stack architecture.
We tune the qubit through several avoided crossings with the resonator and observe the vacuum Rabi splitting between the qubit and the resonator.
The periodicity of these avoided crossings provides the effective inductive coupling between the TSV-integrated bias line and the qubit.
In Figure~\ref{fig:two-stack}b, we measure a separate 2-stack device that incorporates TSV-integrated resonators.
We record complex-valued single-shot measurements for 21,000 preparations each of the lowest two energy states $|0\rangle$ and $|1\rangle$, and fit the distribution of those measurements to a Gaussian function.
The separation and width of the Gaussian fits allow us to extract the separation fidelity.
The $|0\rangle$-prepared single-shot data is also used in Figure~\ref{fig:thermalization}b to extract the qubit temperature.
The data in Figure~\ref{fig:qubit_on_interposer}e are taken from four qubits fabricated on the top side of an interposer chip after TSV fabrication.
All readout resonators are coupled to a common feedline.
Each qubit was sequentially biased to its flux-insensitive point, where the T$_1$ relaxation time was measured 100 times over the course of approximately two hours and then histogrammed.

Measured qubit parameters, including frequencies and T$_1$ relaxation times, are recorded in Table 1.

\section{Data Availability}
The data supporting the findings of this study are available within the paper. The data are available from the authors upon reasonable request and with the permission of our US Government sponsors.

\section{Code Availability}
The code used to process data presented in the study is available from the authors upon reasonable request and with the permission of our US Government sponsors.

\section{Acknowledgements}
We gratefully acknowledge the MIT Lincoln Laboratory design, fabrication, packaging, and testing personnel for valuable technical assistance.
This research was funded by the Office of the Director of National Intelligence (ODNI), Intelligence Advanced Research Projects Activity (IARPA) and by the Assistant Secretary of Defense for Research \& Engineering under Air Force Contract No. FA8721-05-C-0002.
The views and conclusions contained herein are those of the authors and should not be interpreted as necessarily representing the official policies or endorsements, either expressed or implied, of ODNI, IARPA, or the US Government.

\section{Author contributions}
D.R.W.Y., J.M., C.S., J.L.Y., M.C., R.D., D.K.K., A.M., and B.M.N. developed processes and executed the fabrication and 3D integration of the devices.
D.R. and W.W. designed the devices.
M.E.S. performed the cryogenic qubit measurements.
D.R.W.Y., M.E.S., D.R., and W.D.O. wrote the manuscript.
M.E.S., D.R., G.C., A.L.D., E.B.G., and W.W. contributed to qubit design, measurement infrastructure, and data analysis.
D.R., J.L.Y., A.J.K., and W.D.O. provided technical direction.

\section{Competing Interests}
The authors declare no competing interests.

\end{document}